\documentclass[12pt,preprint]{aastex}
\usepackage{natbib}
\slugcomment{submitted to ApJ Letters, \today}

\begin{document}

\title{Satellites of the largest Kuiper belt objects}
\author{M.E. Brown\altaffilmark{1}, 
M.A. van Dam\altaffilmark{2}, A.H. Bouchez\altaffilmark{3},  D. Le Mignant\altaffilmark{2},
R.D.  Campbell\altaffilmark{2}, J.C.Y. Chin\altaffilmark{2}, A. Conrad\altaffilmark{2},
S.K. Hartman\altaffilmark{2}, 
E.M. Johansson\altaffilmark{2}, R.E. Lafon\altaffilmark{2}, D.L. Rabinowitz\altaffilmark{4},
P.J. Stomski, Jr.\altaffilmark{2}, D.M. Summers\altaffilmark{2}, C.A. Trujillo\altaffilmark{5},
P.L. Wizinowich\altaffilmark{2}}
\altaffiltext{1}{Division of Geological and Planetary Sciences, California Institute
of Technology, Pasadena, CA 91125}
\altaffiltext{2}{W.M. Keck Observatory, 65-1120 Mamalahoa Highway, Kamuela, HI 96743}
\altaffiltext{3}{Caltech Optical Observatories, California Institute of Technology, Pasadena, CA 91125}
\altaffiltext{4}{Yale Center for Astronomy and Astrophysics, Yale University, New Haven, CT 06520}
\altaffiltext{5}{Gemini Observatory, 670 North A'ohoku Place, Hilo, HI 96720}
\email{mbrown@caltech.edu}

\begin{abstract}
We have searched the four brightest objects in the Kuiper belt for the presence of satellites
using the newly commissioned
Keck Observatory Laser Guide Star Adaptive Optics system. Satellites are seen around three of
the four objects: Pluto (whose satellite Charon is well-known), 2003 EL61, and 2003 UB313. The
object 2005 FY9, the brightest Kuiper belt object after Pluto, does not have a satellite
detectable within 0.4 arcseconds with a brightness of more than 0.5\% of the primary.
The presence of satellites to 3 of the 4 brightest Kuiper belt objects is inconsistent with
the fraction of satellites in the Kuiper belt at large at the 99.1\% confidence level, suggesting
a different formation mechanism for these largest KBO satellites.
The satellites of 2003 EL61 and 2003 UB313, with fractional brightnesses of 5\% and 2\% of
their primaries, respectively, are significantly fainter relative to their primaries than
other known Kuiper belt object satellites, again pointing to possible differences in their origin.
 \end{abstract}

\section{Introduction}
The discovery and orbital characterization
of satellites to objects in the Kuiper belt has provided a unique
window into the early history of the outer solar system.
The early discovery of Charon \citep{1978AJ.....83.1005C} 
around Pluto and the seemingly high angular
momentum of the Pluto-Charon system led to the
hypothesis that a giant impact was responsible for formation 
of the system \citep{1989ApJ...344L..41M},
suggesting, even before the discovery of the remainder of the Kuiper belt, that
many more objects might exist in the regions beyond Neptune.
It was generally expected that satellites to smaller Kuiper belt objects (KBOs), if they
existed,
would form through the same mechanism and would consequently be on tightly 
bound circular orbits.
The discovery of the first satellite to the smaller Kuiper belt object 1998 WW31 with
a satellite separation of almost three times the separation of Pluto and Charon and with 
a highly eccentric orbit was thus quite a surprise \citep{2002Natur.416..711V}. 
The large semi-major axis of the 1998 WW31 system leads to 
even more specific angular momentum than the Pluto-Charon system. The angular momentum is 
significantly more than can be explained from  impact
formation, leading to the suggestion of a capture origin \citep{2002Natur.420..643G}.
Subsequent discoveries of KBO satellites and the determination of their
orbits have found that most, so far, resemble the 1998 WW31 system \citep{2003EM&P...92..409O, 2004Icar..172..402N,2004AJ....128.2547N}.
The Pluto-Charon system, perhaps because of its size, has been the only 
binary system for which an entirely different formation mechanism has seemed necessary.

With the recent discovery of several KBOs approaching (and even exceeding)
the size of Pluto, we have undertaken a survey of satellites in these systems
in an attempt to determine if the Pluto-Charon system remains unique.
Progress in surveys of newly discovered KBOs has been 
hampered in the past by the long lead time required 
to make observations 
from the  Hubble Space Telescope. Recently, however, a new facility has become available 
for high spatial 
resolution imaging of KBOs from the ground. The W.M. Keck Observatory 
has just 
finished commissioning a Laser Guide Star Adaptive Optics (LGS AO) 
system for use 
with the Keck II telescope which is capable 
of obtaining infrared 
images with a similar 
resolution to that which the HST obtains for visible images \citep{peterw}. 

      An LGS AO system works similarly to natural guide star (NGS) 
adaptive optics 
systems, which have now become standard on large 
telescopes throughout the world, 
in that the aberrations of starlight 
caused by the earth's atmosphere are measured and 
removed by the system. An LGS system, however, 
instead of measuring the aberrations 
using the light from a bright natural star, 
measures aberrations from observations of 589~nm 
laser light resonantly scattered off of sodium 
atoms in a layer at approximately 90 
km altitude in the earth's mesosphere. 
While the use of the laser would ideally  allow the 
LGS AO system to be used to image any location 
visible in the sky, the system has 
several practical limits that can only be 
overcome by having at least one relatively faint 
natural star in moderately close proximity to the 
astronomical target. For current Keck 
LGS AO system performance, near 
diffraction-limited resolution at the K band (2.1 $\mu$m) 
requires a star with magnitude $V<18$ within $\sim$60 
arcseconds of the target \citep{peterw}. Partial 
aberration correction can be obtained 
for a star up to a magnitude fainter \citep{marcos}. 

      Moving KBOs occasionally come close 
enough to bright stars to be able to be 
observed with the LGS AO system, but four 
known objects in the Kuiper belt are bright 
enough that they themselves can be used 
as the natural guide star, and thus they can be 
observed without the time restriction 
imposed by the requirement of a background star. 
We present a Keck LGS AO survey for satellites around these four brightest
known Kuiper belt objects.

\section{Observations}
The four brightest known 
objects in the Kuiper belt, Pluto, 2005 
FY9, 2003 EL61, and 2003 UB313, with V magnitudes of 
14.0, 16.8, 17.5, and 18.8, respectively, 
were all observed with the Keck LGS AO system 
during engineering commissioning in 2005. 
All of the KBOs were observed using an LGS AO setup developed for observing faint 
science targets where the target itself is to be used as the natural star reference. In LGS 
AO, there are quasi-static aberrations resulting from the parallactic elongation of the LGS 
from the perspective of the fast wavefront sensor  that strongly affect the image quality, 
and these aberrations would rotate as the telescope pupil rotates. These aberrations 
are measured at a given pupil rotation angle using a bright nearby NGS.
The pupil angle 
is then kept fixed on the fast wavefront sensor as the telescope is re-pointed to the KBO 
and throughout the observations, ensuring that the point spread function (PSF) remains 
as stable as possible (this step was skipped
for Pluto which is bright enough itself to measure these aberrations). 
The image field, however, rotates about the optical axis as the 
azimuth angle of the telescope changes.  The tip-tilt mirror control loop, which provides 
fast telescope guiding to keep the science target precisely centered, is controlled using a 
quad cell of avalanche of photodiodes guiding on the KBO. The laser is then projected 
at the target and the resonantly scattered laser light is observed by a fast wavefront 
sensor which drives the 349-actuator deformable mirror to correct atmospheric 
aberrations. A final control loop controls the system focus.  The focus of the system 
changes slowly due to changes in the structure of the mesospheric sodium layer. These 
focus changes are measured using a low-bandwidth wavefront sensor that looks at the 
KBO, and the measurement is used to drive the position of the fast wavefront sensor to 
be optically conjugate to the height of the sodium layer. 

Once the LGS AO system is set up, observing proceeds identically to standard 
IR observing procedures. All the KBOs were observed 
through a K' filter (1.948-2.299 $\mu$m) with the NIRC2 imager. The brightest three KBOs 
were imaged using the 9.9 milliarcsec plate scale camera while 2003 UB313 was 
imaged using the 39.7 milliarcsec plate scale camera since it is fainter and the 
correction is consequently worse.   

For Pluto, three 10-second exposures were taken at dither positions separated by 2 arcsec
on the detector, for a total integration time of 30 seconds. 
For 2005 FY9 and 2003 EL61 six 60-second exposures were taken at each of four dither 
positions, for a total integration time of 720 seconds. For 2003 UB313 six 60-second exposures
were obtained at each of four dither positions, for a total integration time of 1440 seconds.
The images were corrected for sky and instrumental background by subtracting 
the median of the images in each dither pattern. They were then flat-fielded using 
twilight sky flats and known bad pixels were interpolated over. The individual images 
were then combined, correcting for rotation of the image with time, by shifting to a 
common center by cross-correlation. 

Figure 1 shows Keck LGS AO images of the four brightest objects in the Kuiper belt. 
The image of Pluto clearly shows its known satellite, Charon. Faint point sources are also 
seen near 2003 EL61 and 2003 UB313, while none is seen around 2005 FY9. These 
faint point sources are clearly not background stars or galaxies, as they move across the 
sky with the same motion as the primaries. They also cannot be artifacts of the LGS AO 
system, as the PSF of the LGS AO system rotates in the image plane during the course of
the observations and thus any PSF artifact would be smeared into an arc.
We thus conclude 
that Pluto, 2003 EL61, and 2003 UB313 are circled by satellites. 
No satellite is seen in the vicinity of 2005 FY9. 
Experiments with embedding artificial point source into
the 2005 FY9 field suggest that a satellite with a fractional brightness of 0.5\% that of 2005 
FY9 would have been seen at a separation of 0.4 arcseconds or greater. Table 1 summarizes 
the detections or upper limits of the satellites for these four brightest objects in the 
Kuiper belt. 

      After the initial Keck LGS AO discovery of the satellite to 2003 EL61,
five observations over the course of six months were used to 
determine that the mass of the 2003 EL61 system is 4.2 x 1021 kg or 32\% that of Pluto
\citep{me}. 
With only a single observation of the satellite of 2003 UB313 we cannot yet measure or 
constrain the mass of 2003 UB313, but we can estimate likely orbital parameters to aid 
further study. If the satellite is on a circular (like Charon) or near-circular (like the 2003 
EL61 satellite) orbit with a random orientation, then at any random point in time it is 
50\% likely to be at a separation within 14\% of its semi-major axis. 
An early search for a satellite to 2003 UB313 using NACO at the 
VLT  and a chance appulse with a bright natural guide star would have detected the satellite
out to a separation of approximately 0.4 arcseconds, but saw no satellite \citep{vivanov}, 
suggesting that
the satellite does not spend all of its time at the current separation.
If the semimajor axis 
is 14\% greater than  the current separation, and if 2003 UB313 has the size estimated by 
assuming an albedo and  density similar to Pluto's \citep{me2}, the satellite will have an orbital 
period of approximately two weeks. Observations over the coming season will allow an 
accurate determination of the mass of this planetary-sized body.

\section{Discussion}

Three out of four of the brightest known objects in the Kuiper belt have satellites. The most
extensive Hubble Space Telescope 
survey of the general KBO population to date found 9 satellites out of 81
observations\citep{2005DPS....37.5616S}. 
The probability that these two populations have the same satellite fraction, $f_s$,
can be calculated from simple binomial probability theory. Given 9 satellites out of 81 observed
objects, the
probability distribution, $P[f_s]$ for $f_s$ can be calculated as 
$$
P[f_s]={{_9C_{81}f_s^9(1-f_s)^{72} } \over {\int_0^1 {_9C_{81}}f'^9(1-f')^{72} df'}},
$$
where $_9C_{81}$ is the number of unique ways to chose 9 objects out of a sample of 81,
calculated as $81!/(81-9)!9!$. The probability, $P_{3+}$ that 3 or more out of 4 objects observed would
then have a satellite is given by
$$
P_{3+}=\int_0^1 P[f_s] ({_3C_4}f_s^3(1-f)+ {_4C_4}f_s^4) df_s.
$$
For the current sample, $P_{3+}$, the probability that the two populations have the same value
of $f_s$, is equal to 0.9\%. Thus, even with the very small sample involved,
the result that the large satellites and the smaller ones do not have the same probability of 
having a satellite is significant.

While the survey of \citet{2005DPS....37.5616S} has not yet 
published detection limits, it is likely that 
for many of the objects observed in this survey, 
faint satellites like those of 2003 EL61 and 2003 UB313 would not have been detected.
 Thus the difference
in fractional abundance between the two populations could simply be due to the greater 
relative depth of the LGS AO survey. A smaller but deeper HST program surveyed 19 satellites to
a depth sufficient to have detected satellites with a fractional brightness of 1\% within 0.3
arcseconds of the primary (Trujillo and Brown, in prep). This survey detected two satellites within these limits.
Using the above binomial calculation, the probability that a sample of 3 
or more out of 4 and 2 out of 19 would be drawn from the same fraction is only 1.8\%. Again, 
even with the small number of objects surveyed the difference between the two populations is
significant. It thus appears that the overabundance of satellites to the brightest KBOs 
is intrinsic to this population rather than a function of survey limits.

The satellites of 2003 EL61 and 2003 UB313 are much fainter compared to their primaries than
any other known satellites.
Neither of these satellites appears likely to have 
formed from the process of dynamical-friction aided capture thought to have occurred 
for many smaller Kuiper belt objects \citep{2002Natur.420..643G} 
as this process requires that small bodies drain 
energy from the larger bodies to aid the capture. For bodies as faint as these satellites,
dynamical friction would be essentially inoperable. Numerical simulations of a 
collisional origin for the Pluto-Charon system have been explored in detail
\citep{2005Sci...307..546C}, and many 
of the potential system outcomes after an impact contain satellites with a relative sizes 
similar to the 2003 UB313 and 2003 EL61 satellites. The simulated formation of these 
smaller satellites differs from the simulated creation of the Pluto-Charon system in that 
the large size and angular momentum of Charon are best produced by intact formation 
following the impact, while smaller sized objects are formed in accretion disks similar 
to that thought to have formed the Moon after an impact on the Earth. Formation in a 
disk has been shown to lead to a more-rapidly spinning primary \citep{2005Sci...307..546C}, which could 
also explain 
the unusually rapid rotation of 2003 EL61 \citep{david}. Nothing is currently known about the 
rotation state 2003 UB313, but the small secondary to 2003 UB313 might suggest a 
similarly rapid rotator. 
While simulations suggest that a giant impact with special geometry is required 
to explain the large mass fraction of Charon, smaller satellites appear to be able to be 
formed around Pluto-scale KBOs with a much wider range of impact geometries 
\citep{2005Sci...307..546C}.  
While once Pluto appeared unique in the outer solar system in terms of size and satellite 
formation mechanism, it now appears to be one of a family of similar-sized objects with 
perhaps similar collisional histories and a range of satellite outcomes.

 {\it Acknowledgments:}
Data presented herein were obtained at the W.M. Keck Observatory,
which is operated as a scientific partnership among the California
Institute of Technology, the Universities of California, and the National
Aeronautics and Space Administration. The observatory was made possible
by the generous financial support of the W.M. Keck Foundation.
The authors wish to recognize and acknowledge the very significant 
cultural role and reverence that the summit of Mauna Kea has always 
had within the indigenous Hawaiian community.  We are most fortunate 
to have the opportunity to conduct observations from this mountain.
We are grateful to Chuck Sorenson, James Lyke, Cynthia Wilburn and Christine 
Melcher for their assistance with the observations, to
Sniffen Joseph, David Collier, Nick Jordan, Teresa Kim-Pedro, and Michael
Ruff for enduring long cold nights outside making sure the laser did not
shoot down any airplanes, 
 and to Kris Barkume and Emily Schaller for 
discussions of Kuiper belt object satellite formation and populations.

\newpage

\begin{figure}
\plotone{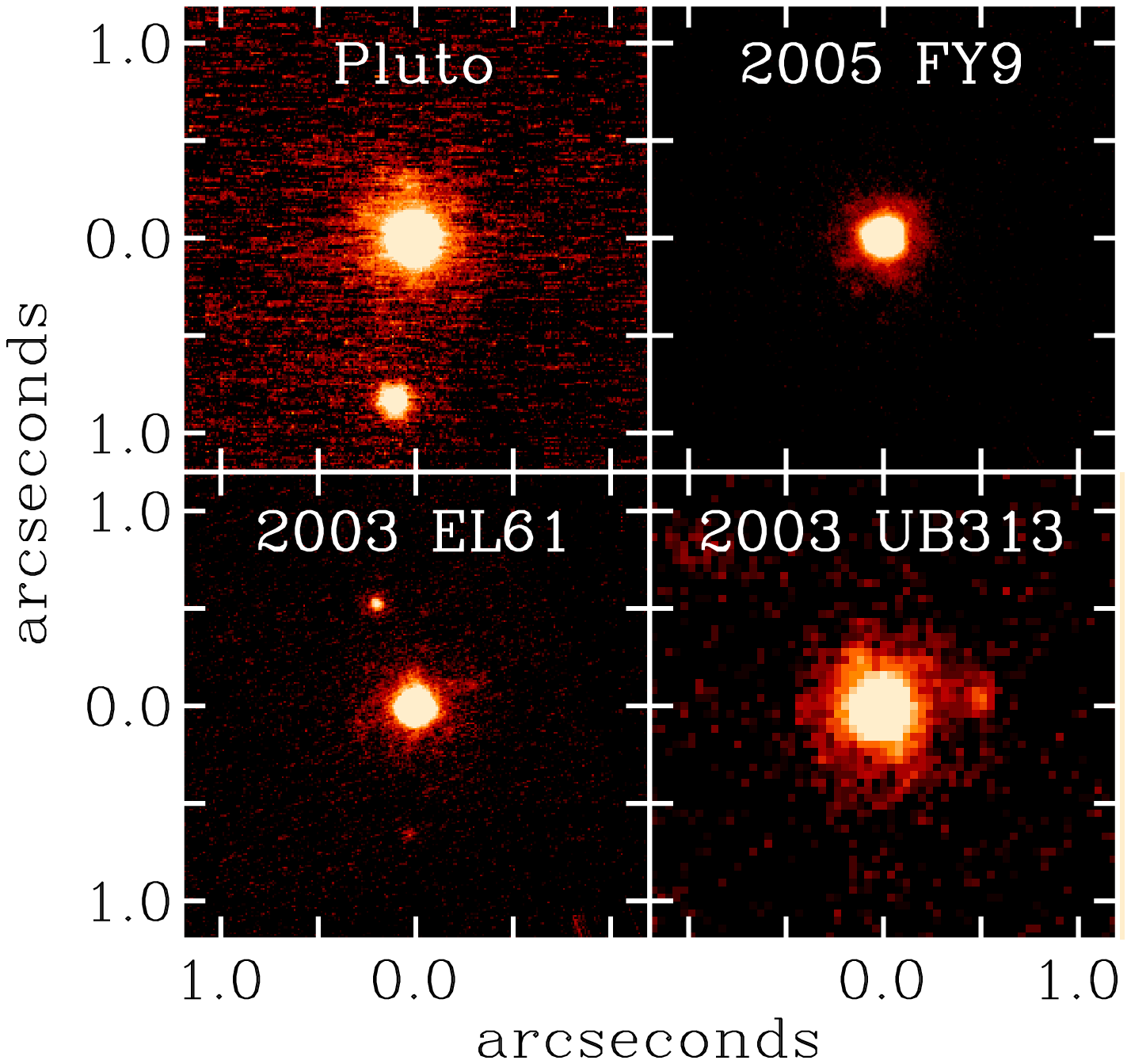}
\caption{Images of the four brightest Kuiper belt objects from the Keck Observatory Laser Guide Star Adaptive Optics system. All images are identically scaled logarithmically to the brightest point
of the Kuiper belt object and oriented with north up. Satellites are seen clearly Pluto (directly
below), 2003 EL61 (above and left; the faint source directly below is a trailed background star),
and 2003 UB313 (directly right).
}
\end{figure} 

\begin{table}
\begin{center}
\caption{Parameters of satellites of the largest Kuiper belt objects}
\begin{tabular}{lcccc}
\tableline\tableline
 & Pluto& 2005 FY9& 2003 EL61& 2003 UB313\\
\tableline
Observing date (2005)& 11 September & 28 May & 30 June& 10 September\\
V magnitude&14.0&16.8&17.5& 18.8\\
FWHM (arcseconds)\tablenotemark{a} &0.058 &0.068& 0.063& 0.120\\
Strehl ratio\tablenotemark{b} &0.37& 0.20& 0.18& 0.10\\
Fractional brightness of satellite& 0.19& $<0.005$ &0.05 0&.02\\
Apparent semi-major axis (arcseconds) &0.87 &$>0.4$& 1.3& 0.53\tablenotemark{c}\\
True semi-major axis  (km)     & 19640&-&49500&36000\tablenotemark{c}\\
Orbital period (days)             &6.4&-&49&$\sim$14\tablenotemark{d}\\
\tableline
\tablenotetext{a}{ Full width at half maximum of the image of the primary, showing the near diffraction-limited 
performance.}
\tablenotetext{b}{Strehl is a measure of the peak intensity of the image compared to the theoretical expectation 
for a diffraction-limited image.}
\tablenotetext{c}{Observed separations only for 2003 UB313.}
\tablenotetext{d}{Crude estimate. See text.}
\end{tabular}
\end{center}

\end{table}

\end{document}